\documentclass[epj]{svjour}
\usepackage[dvips]{graphics}
\usepackage{psfrag}

\begin{document}

\title{Compaction of a granular material under cyclic shear}

\author{M. Nicolas, P. Duru \and O. Pouliquen}  
 
\institute{Groupe \'Ecoulements de Particules\\ IUSTI Technop\^ole de 
Ch\^ateau-Gombert, 5 rue Fermi 13453 Marseille cedex 13 France \\
email: maxime@iusti.univ-mrs.fr}
 
\date{Received: date / Revised version: date}
 
\abstract{
In this paper we present experimental results concerning the compaction of a granular assembly of spheres under periodic shear deformation.  The dynamic of the system is slow and continuous when the amplitude of the shear is constant, but exhibits rapid evolution of the volume fraction when a sudden change in shear amplitude is imposed. This rapid response is shown to be to be uncorrelated with the slow compaction process. 
\PACS{
      {45.70.Cc}{Granular compaction}   \and
      {61.50.-f}{Crystalline state}
     } 
} 
 
\maketitle

\section{Introduction}

Powder compaction has recently attracted the attention of physicists 
as the prototype of a disordered system without thermal 
fluctuations.  For millimetric grains, thermal energy is negligible 
compared to gravitational energy. Once poured in a box, the grain 
packing is trapped in a meta\-stable configuration and stay in this 
state unless an external excitation such as vibrations is imposed.  
During such a process, the particles assembly exhibits a slow and 
complex evolution towards a more compact configuration. This 
evolution has been studied experimentally by Knigth \textit{et al} 
\cite{knight95} and Nowak \textit{et al} \cite{nowak98} in the case 
of a vertically vibrated packing of spheres.

Their experiment consisted in a vertical cylinder full of 
monodispersed beads submitted to successive distinct vertical taps of 
controlled acceleration.  The mean volume fraction was recorded after 
each tap.  The experiments were first performed with taps of constant 
amplitude \cite{knight95}. They have shown that the increase in 
volume fraction is a very slow process well fitted by the inverse of 
a logarithm of the number of taps. The more energetic taps are the 
more 
efficient for compaction.  Nowak \textit{et al} \cite{nowak98} have 
studied the compaction under taps of variable amplitude, and showed 
that irreversible processes occur during the compaction. Starting 
from a loose packing, the evolution of the volume fraction is not the 
same when increasing the amplitude of vibration as when decreasing. 
The first branch is irreversible, whereas the second is reversible. 
The highest volume fraction is obtained by first increasing the tap 
acceleration then decreasing it back to zero.
Recently, Josserand \textit{et al} \cite{josserand00} have 
investigated the evolution of the packing to sudden change of tap 
amplitude. They showed that the response of the system depends on the 
history of compaction.

These experimental results have motivated numbers of theoretical and 
numerical studies trying to find a minimal model which could exhibit 
qualitatively similar features.  Most of the models suggest that the 
rearrangement of particles in a more compact state 
needs more and more cooperative reorganization as the volume fraction 
increases.  Absorption-desorption model 
\cite{nowak98,jin94,talbot99}, geometrically frustrated 
lattice model \cite{coniglio96,nicodemi97,nicodemi99}, random walk model \cite{luding00} or excluded 
volume model have been proposed \cite{edwards,boutreux97}.
Most of these approaches aspire to describe the vertically vibrated 
system. To our knowledge the question of how the compaction dynamics 
depends on the method of excitation has not yet been raised. In 
absence of thermal fluctuations,
one can legitimately wonder how the compaction dynamic depends on the 
way the system is excited.

An alternative way of excitation is horizontal vibration. In a 
previous experiment we have shown that very compact and crystalline 
packings of monodispersed beads can be obtained this way 
\cite{pouliquen97} whereas vertical vibrations only yield random 
close packings.  However, under horizontal vibration, only the top 
layers of the packing are affected and indeed submitted to a periodic 
horizontal shear. 
The goal of this paper is the study of the response to shear of the whole 
packing. 
A parallelepiped box full 
of beads is submitted to a horizontal shear through the periodic 
motion of two parallel walls.  
The resulting compaction is analyzed in detail in this paper.

Deformation of granular material under cyclic solicitation has been 
extensively studied in soil mechanics \cite{pande80}.  Cyclic loading 
in a triaxial 
test shows a compaction but no systematic study of the dynamic of 
compaction for long time behavior has been investigated.  Cyclic 
shear is also evoked by Scott \textit{et al} \cite{scott64} in a 
short note where they show that this deformation applied to 
monodispersed beads packing could lead to a partially ordered 
structure.   A shear apparatus has been also developed by Bridgwater 
and Scott \cite{scott76,cooke78} 
in order to study the segregation dynamic during shear.  Cyclic shear 
has been recently applied to 
thermally activated systems such as assembly of spherical colloidal 
particles.  A periodic shear deformation applied to a colloidal hard 
spheres glass yields the formation of ordered regions 
\cite{haw98a,haw98b}.  However, both the thermal excitation and the 
imposed deformation are in 
this case 
present.  The experiments presented in this paper can be seen as the 
analog of the colloidal experiments at zero temperature.  The paper 
is presented as follows.  The experimental setup is described in 
section 
\ref{s:expsetup}. In section \ref{s:constant} the compaction under 
constant shear amplitude is 
investigated and the crystalisation process and the slow 
relaxation are discussed.  The section \ref{s:memory} reports the 
study of the response 
of the system to sudden change in the shear amplitude.  
Finally we discuss in section \ref{s:discussion} the analogies and 
differences between compaction under cyclic shear and vertically 
vibrated 
system before giving concluding remarks in section \ref{s:conclusion}.

\section{Experimental setup}
\label{s:expsetup}

\begin{figure}
	\psfrag{h}{$h$}
	\psfrag{D}{$D$}
	\psfrag{position}{position sensor}
	\psfrag{q}{\large$\theta$\normalsize}
	\psfrag{cable}{cable}
	\psfrag{hinge}{hinge}
	\psfrag{bottom}{bottom plate}
	\psfrag{glass}{glass}
	\psfrag{plate}{plate}
	\resizebox{0.45\textwidth}{!}{\includegraphics{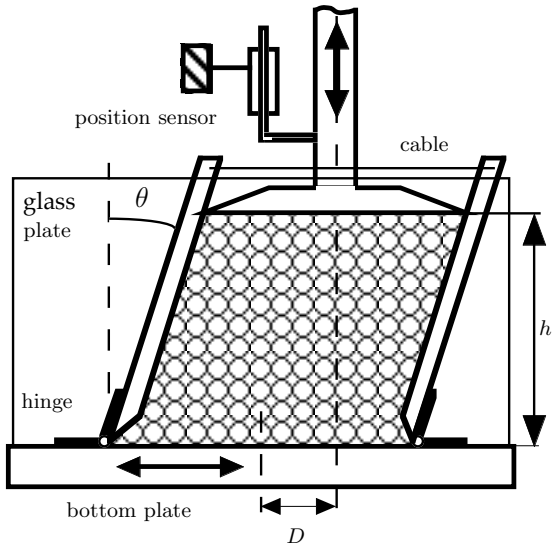}}
	\caption{Schematic of the shear cell and position sensor. $D$ is the 
horizontal displacement of the bottom plate and $h$ is the height of 
the packing. Angle of shear is $\theta=\arctan(D/h)$.}
	\label{fig21}
\end{figure}

The experimental setup is sketched in Fig. \ref{fig21}.  The shear 
cell was 
parallelepiped, and the volume occupied by the beads was typically 
10.5 cm 
high, 7.9 cm wide and 10.2 cm deep.   The bottom of the cell was a 
flat 
PVC plate attached to a 
horizontal linear displacement device driven by a stepper motor.  The 
two mobile lateral walls were aluminum plates linked 
with two hinges to the bottom plate.  These plates were covered with 
a 
thin plastic sheet to insulate the packing from metallic dust due to 
particle/wall friction.  The front and back sides were glass plates 
fixed on the bottom plate.

The granular packing was confined on the top by a rectangular plate 
mounted on a vertical displacement rail.  This plate was independent 
from the rest of the shear cell and was free to move vertically only, 
no
horizontal displacement being allowed.  The periodic shear 
deformation was obtained by imposing a periodic horizontal 
displacement to the bottom plate, the two mobile lateral walls being 
constrained by two cables to remain in contact with the top plate.

The granular material we used were spherical glass beads.  
Experiments were performed with either monodispersed  or bidispersed 
set of beads.  For experiments with monodisperse particles, beads 
were $2.97\pm 0.06$ mm in diameter with a density of $2.52\pm 
0.02$~g.cm$^{-3}$. The cell was filled with 1450~g of particles. The 
bidispersed set of particles was constituted of $725$~g of the 
$3$~mm beads and $725$~g of $2$~mm beads of the same density.
In order to prevent the surface deterioration 
of the beads during long-time experiments, beads were coated with 
silicon oil Rhodorsil 47V100 (viscosity hundred times viscosity of 
water).  Note that the same quantitative results were observed 
without 
lubrication, but lifetime of the beads was in this case quite 
shorter.

We measured the
 volume fraction during the compaction process by recording 
the vertical position of the top 
plate, which  went down (resp.  up) when 
compaction (resp.  dilatation) occurred. Its position was accurately 
measured by a linear position sensor 
Novotechnik T50 (potentiometric transducer). The position 
resolution was $3\,10^{-3}$~cm corresponding to a $2\,10^{-4}$ 
resolution in volume fraction. Both the data acquisition and the 
horizontal displacement of the cell were controlled by the same PC 
computer. Data could be recorded continuously during the shear, or 
once 
every cycle for the study of the long time evolution. 

The experimental procedure was the following. The top plate was 
removed, the mobile side walls were put the vertical position and the 
particles were poured into the cell through a hopper.  
The mean initial 
volume fraction of the packing obtained by this procedure 
$\phi_0=0.592\pm0.008$.  
The top plate was then slowly put into contact with the 
packing.  Once the initial random packing was ready, the periodic 
shear 
deformation was imposed.  The lateral plates were inclined to an 
angle 
$+\theta$, followed by an inclination to an angle $-\theta$.  
The plates were then put back in the vertical position and the volume 
fraction was recorded.  A new cycle of shear was then applied and so 
on. The volume fraction could also be continuously recorded.

During the experiments, we have observed that the inclination angle 
$\theta$ was the only pertinent control
parameter. Change in the velocity of the bottom plate (from $1.9$ to 
$3.8$~cm/s) 
or change in the weight exerted by the top plate on the packing (from 
$1.8$ to $3.8$~kg) did not affect any of the measurements.
We also performed experiments using asymmetric motion of the 
lateral plates: the plates were oscillating between 
$\theta_{medium}+\theta$ and $\theta_{medium}-\theta$. In the range 
accessible by our set-up ($\theta < 12^\circ$) no influence of 
the medium position has been observed. The compaction process was 
then only controlled by the  amplitude of 
the cyclic shear. 

\section{Cyclic shear of constant amplitude}
\label{s:constant}

\begin{figure}
	\psfrag{p}{$\phi$}
	\psfrag{n}{$n$}
	\resizebox{0.45\textwidth}{!}{\includegraphics{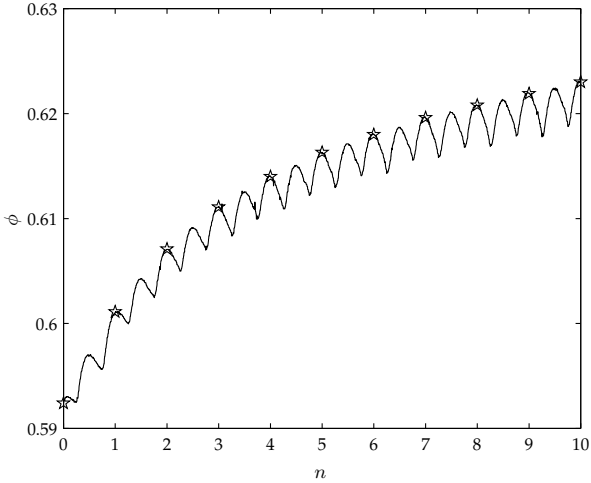}}
	\caption{Continuous evolution of volume fraction as a function of number of cycles; $\theta=5.4\,^\circ$.}
	\label{fig31}
\end{figure}

We have first studied the evolution of a packing submitted 
to a periodic shear of constant amplitude $\theta$. 
A typical evolution 
of the packing volume fraction is presented in Fig. \ref{fig31}.  In 
this 
figure the volume fraction $\phi$ was recorded with a resolution of 
325 samples/cycle.  This plot clearly 
shows a slow and continuous evolution from the initial random packing 
towards higher volume fraction.  Superposed to this slow evolution, a 
quasi-periodic oscillation is observed, corresponding to the 
dilatation occurring during one cycle: the packing is 
denser in the vertical position and alternatively dilates when the 
walls 
are inclined at $\theta$ and $-\theta$.  This behavior is 
also observed when the medium position is not the vertical.  In this 
case a local maximum of the volume fraction is measured at the medium 
position $\theta_{medium}$ whereas a local minimum is observed when 
the walls are inclined at $\theta_{medium}+\theta$ and 
$\theta_{medium}-\theta$. 

\begin{figure}
	\psfrag{q}{$\theta\ (^\circ)$}
	\psfrag{p}{$\phi$}
	\psfrag{n}{$n$}
	\resizebox{0.45\textwidth}{!}{\includegraphics{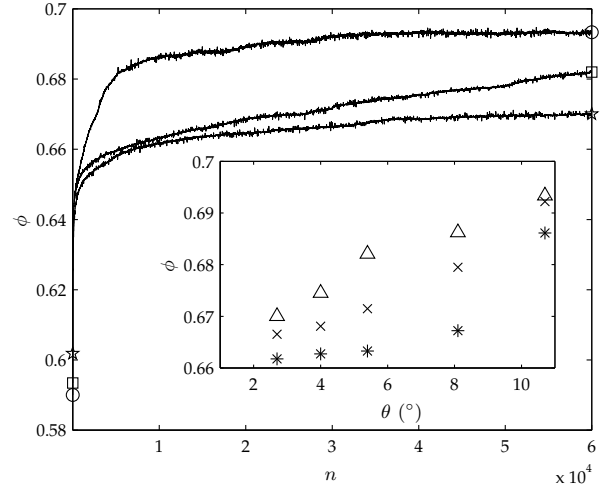}}
	\caption{Compaction under different shear angles : 
$\theta=2.7^\circ$ (star), $\theta=5.4^\circ$ (square), 
$\theta=10.7^\circ$ (circle). Insert: volume fraction of the packing 
versus shear angle for increasing number of cycles: $n=10^4$ cycles 
(asterisks), $n=3\,10^4$ cycles (crosses), $n=6\, 10^4$ cycles 
(triangles).}
	\label{fig32}
\end{figure}

In this paper we do not study the dilatancy properties in detail but 
rather focus on the long time evolution of the packing volume 
fraction. For this purpose, the volume fraction is recorded only once 
a cycle when the walls are brought back to the vertical position 
(stars in Fig. \ref{fig31}).  A monotonous curve is then obtained, 
representing the evolution of the volume fraction $\phi$ as a 
function of the number of cycles $n$.  The results for three 
different shear amplitudes $\theta=2.7\,^\circ,\ 5.4\,^\circ,\ 
10.7\,^\circ$ are presented in Fig. \ref{fig32} for the case of 
$3$~mm beads.  

The first remark is that the compaction process is a very slow 
process.  After $6\,10^4$ cycles (one week experiment), 
the volume fraction still slowly increases. This slow compaction 
appears to be more efficient for large shear amplitude than for small 
shear amplitude as shown by insert in Fig. \ref{fig32} representing 
the volume fraction as a function of the shear amplitude $\theta$ 
after $10^4$ (asterisks), $3\,10^4$ (crosses), and $6\,10^4$ cycles 
(triangles). 

\begin{figure}
	\psfrag{p}{$\phi$}
	\psfrag{n}{$n$}
\resizebox{0.45\textwidth}{!}{\includegraphics{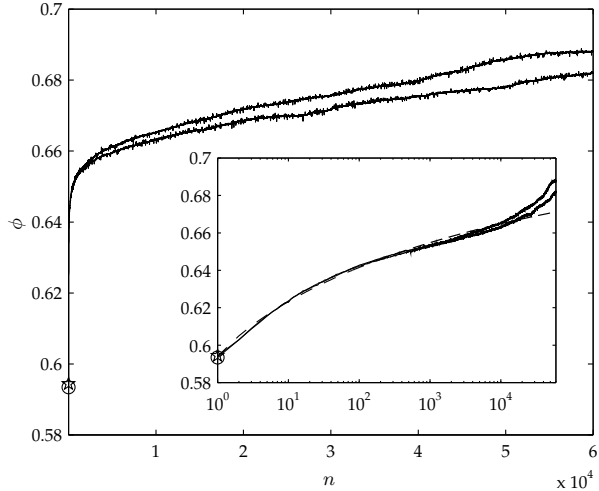}}
\caption{Compaction curves for $\theta=5.4\,^\circ$ for two different 
runs. Insert: semi-logarithmic scale. Dotted line is the inverse 
logarithmic fit 
$\phi=\phi_{\infty}-(\phi_{\infty}-\phi_0)/[1+B\ln(1+n/\tau)]$ with 
$\phi_{\infty}=0.74$, $\phi_0=0.594$, $B=0.1$ and $\tau=0.8$.}
\label{fig33}
\end{figure}

Although curves of Fig. \ref{fig32} present very small fluctuations 
for an individual run, the final volume fraction can vary 
significantly (approximately $10 \%$) from one experiment to another 
carried out at the same shear angle. For example, Fig. \ref{fig33} 
shows two compaction curves obtained for $\theta=5.4\,^\circ$ 
starting from almost the same initial volume fraction 
($0.594\pm0.0008$) prepared by the same procedure. Evolution of the 
packing thus seems to be rather sensitive to the initial random 
structure of the grains assembly.
Insert of Fig. \ref{fig33} presents the same curves in semi-logarithmic 
scale. We have tried to fit our experimental results by  the inverse 
logarithmic function proposed by Knight \textit{et al} for the 
compaction under vertical taps (Eq. 3 of \cite{knight95}). However, 
no reasonable agreement with both the short and long time behavior 
could be found (Fig. \ref{fig33}).  Other proposed fits like the two 
exponential function \cite{barker93} or the stretched exponential 
(\cite{knight95} and references therein) do not provide more 
convincing agreements. 

\begin{figure}
	\resizebox{0.45\textwidth}{!}{\includegraphics{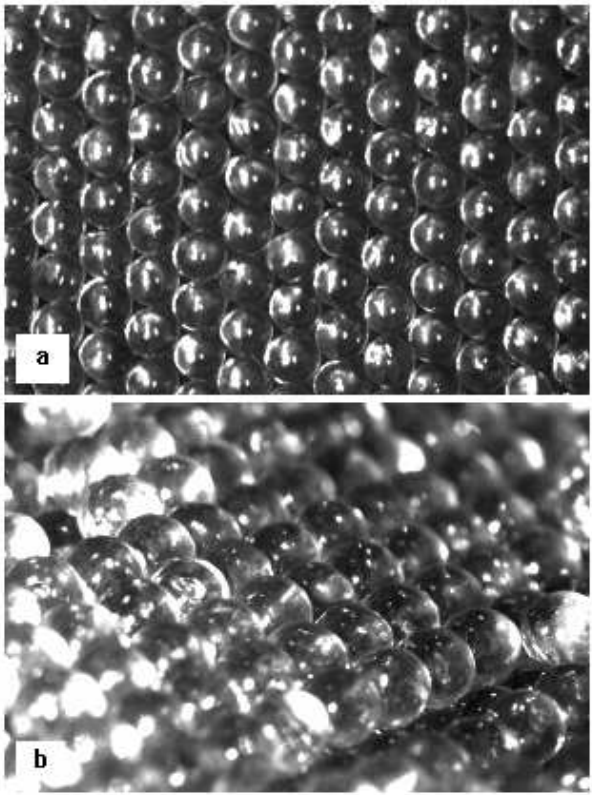}}
	\caption{Photographs of the 3 mm beads packing. (a) front view of 
the packing. (b) Ordered structure in the bulk of the packing (upper 
layers of the packing were removed).}
	\label{fig34}
\end{figure}

A major difference between the periodic shear compaction and 
the vertically vibrated system is 
the large  volume fraction up to $0.693$ that can be attained by the 
former. 
Such a value obtained 
for monodispersed particles is much larger than the volume fraction 
of a 
random 
close packing, which is $0.63$, according 
to Scott and Kilgour \cite{scott69}.
This means that crystalline arrangements are created during the 
compaction, which perhaps explains why the $\log t$ behavior is not observed in
our system. 
The crystal structure of the packing is indeed observed at the wall 
of 
the cell as shown in Fig. \ref{fig34}a. Moreover we checked that 
crystallisation is also present in the bulk by carefully removing the 
particles layer after layer at the end of an experiment. The 
picture in Fig. \ref{fig34}b shows a structure observed in the bulk. The orientation of the crystal
observed in the bulk is not always parallel to the walls, suggesting that the order is not only
wall-induced but nucleates and grows in the bulk. 

In contrast, no
crystalline alloy structure has been observed with 
 the bidispersed material made of 
mixture of 2 and 3 mm beads. 
However the evolution of the volume 
fraction as a function of number of cycles and shear amplitude 
presents the same trend using mono or bidispersed material. 
In this case, the maximum volume fraction obtained for 
$\theta=10.7\,^\circ$ after $6\,10^4$ cycles is 
$0.65$ to be compared with $0.693$ for the monodispersed packing.

\section{Response to a change in shear amplitude}
\label{s:memory}

\begin{figure}
	\psfrag{p}{$\phi$}
	\psfrag{n}{$n$}
	\psfrag{A}{A}
	\psfrag{B}{B}
\resizebox{0.45\textwidth}{!}{\includegraphics{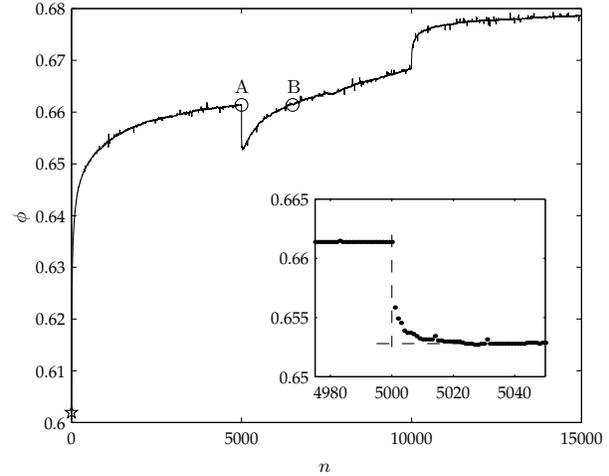}}
\caption{Example of angle variation effect during the compaction 
process. Insert shows a close-up of the first jump (point A).}
\label{fig41}
\end{figure}

After the study of compaction under constant shear angle, we 
investigated the effects of a sudden change of angle.
A periodic shear with inclination angle $\theta_1$ is first imposed 
to a random packing, and at a given time, the shear amplitude is 
suddenly changed to another value $\theta_2=\theta_1+\Delta\theta$. A 
typical experiment is presented in Fig. 
\ref{fig41}. 
Starting with a shear amplitude $\theta_1=2.7\,^\circ$, the angle is 
switched to $\theta_2=10.7\,^\circ$ at $n=5000$, and then switched 
back to $\theta_1$ at $n=10^4$. As can be seen, increasing the shear 
angle at $n=5000$ produces a rapid fall of volume fraction $\phi$ 
followed by a slow and continuous increase. When shear angle is 
decreased back at $n=10^4$, a rapid increase of $\phi$ occurs, 
followed by a slower one. 

Such a response is non-trivial. It first shows that memory effects 
exist in the compaction process. Points A and B in Fig. \ref{fig41} 
correspond to packings having the same volume fraction. However, 
their responses to the same shear amplitude $\theta=10.7\,^\circ$ are 
different: packing A become looser whereas packing B pursues its 
compaction. This means that knowing the volume fraction of the 
packing is not sufficient to predict the evolution of the system.
Such memory effects have been recently put in evidence in vertically 
shaken granular packings \cite{josserand00}.

Another important observation is the rapid character of the packing 
response. 
Whereas the compaction from initial random state is a slow process 
(Sec. \ref{s:constant}), a sudden variation of shear amplitude 
induces a rapid change of volume fraction, which can be regarded 
as a discontinuity or a jump. The amplitude $\Delta\phi$ of this jump 
can be accurately 
defined and measured with the help of the tangent lines as shown on 
the insert of Fig. \ref{fig41}. This definition is also valid for 
positive jumps ($n=10^4$ on Fig. \ref{fig41}).
We have systematically studied how the volume fraction jumps depend 
on the different parameters. 
  
\begin{figure}
 	\psfrag{dp}{$\Delta\phi$}
	\psfrag{time}{Age of packing}
 	\resizebox{0.45\textwidth}{!}{\includegraphics{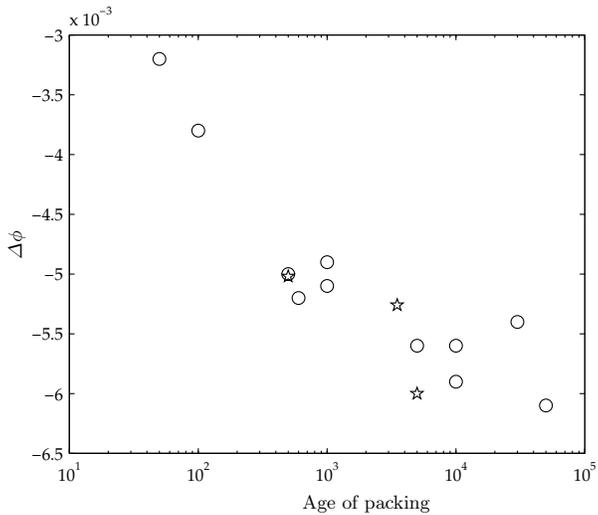}}
 	\caption{Jump amplitude versus age of packing. 
$\Delta\theta=5.4\,^\circ$ (circles), $\Delta\theta=-5.4\,^\circ$ 
(stars).}
 	\label{fig43}
\end{figure}

First, we have investigated how the discontinuity $\Delta\phi$ is 
influenced by the age of the packing, namely the time elapsed before 
the angle change is applied. Fig. \ref{fig43} shows the response 
$\Delta\phi$ to the same angle change $\Delta\theta=5.4\,^\circ$ as a 
function of the elapsed time. The jump $|\Delta\phi|$ increases with 
age and eventually saturates for $n>2000$. The same trend is observed 
for positive jumps ($\Delta\theta=-5.4\,^\circ$). After 2000 cycles, 
the response does not depend on the age anymore.

Secondly, we have observed that the initial angle does not play any 
role. Changing the shear amplitude from $\theta_1$ to 
$\theta_1+\Delta\theta$ yield a response $\Delta\phi$ in volume 
fraction independent of $\theta_1$. \textit{In fine}, the response of the 
system for sufficiently aged packings only depends on the shear change
 $\Delta\theta$.

\begin{figure}
	\psfrag{dp}{$\Delta\phi$}
	\psfrag{dq}{$\Delta\theta\ (^\circ)$}
	\resizebox{0.45\textwidth}{!}{\includegraphics{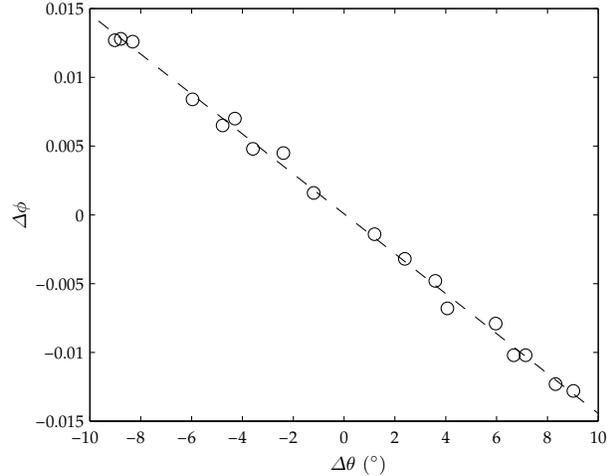}}
	\caption{Relation between volume fraction jumps and angle change for 
aged packings ($n>3\,10^4$).}
	\label{fig42}
\end{figure}

Fig. \ref{fig42} summarizes jump measurements carried out for the 
number of cycles $n>3\,10^4$. Data points align along a straight line 
crossing the origin:
\begin{equation}
	\Delta\phi=-\alpha\Delta\theta
	\label{e:saut_lin}
\end{equation}
with $\alpha=1.45\,10^{-3}$ and $\Delta\theta$ in degrees. 
In conclusion, the response of the packing to sudden change amplitude 
appears to be simpler than first thought. 
The rapid variation of the volume fraction induced by the change is 
simply proportional and opposite to the angle change. 
 
\section{Discussion}
\label{s:discussion}

Two kinds of evolution are observed in the system of cyclic sheared 
packing: a slow and continuous compaction when the shear amplitude is 
constant and a rapid response when a change in shear amplitude is 
imposed. The striking feature is that the rapid dynamic is 
uncorrelated with the slow compaction: the variation of the volume 
fraction $\Delta\phi$ induced by a sudden change  $\Delta\theta$ in 
shear amplitude is independent of the state $\phi$ of the system, and 
depends linearly on $\Delta\theta$. This observation suggests to 
split the evolution of the volume 
fraction in two terms:
\begin{equation}
	\phi(n)=\phi_{slow}(n) - \alpha \theta.
	\label{e:evol}
\end{equation}
The first term $\phi_{slow}(n)$ corresponds to a slow and 
\textit{continuous} 
evolution and the second term is proportional to angle $\theta$.
This description is compatible with Fig. \ref{fig42}: when a sudden 
change in $\theta$ is imposed, one observes a corresponding jump 
$\alpha \Delta \theta$ in volume fraction, $\phi_{slow}$ being a 
continuous function. 

One way of interpreting these results is to say that the 
periodic shear deformation plays two roles.  On one hand, it allows 
for 
the particles to slowly and continuously rearrange to form a
more compact packing. It thus profoundly affects the 
structure 
of the packing in an irreversible way, by introducing more and more 
order.  This contribution is $\phi_{slow}$. 
On the other hand, it introduces an additional disorder superimposed to the ordered structure, represented in Eq. \ref{e:evol} by  
the negative contribution $-\alpha \theta$. These two contributions are uncorrelated.
In this framework, the slow and rapid dynamics would then correspond 
to changes in the packing structure that are of complete different 
nature. 

However, such interpretation remains speculative as long as we are 
not 
able to relate the macroscopic behavior of the volume 
fraction to local rearrangement of the particles.  An important 
question is the discrimination between the 
rearrangements involved in the slow compaction process from 
those 
involved during the rapid compaction or dilatation subsequent to a 
sudden change in shear amplitude. Some numerical simulations 
suggest the existence of collective motion of cluster and of 
individual 
motion of particles which could perhaps be related to the slow and 
rapid dynamics \cite{barker92,barker93,head98}. 

\begin{figure}
	\psfrag{p}{$\phi$}
	\psfrag{q}{$\theta\ (^\circ)$}
	\resizebox{0.45\textwidth}{!}{\includegraphics{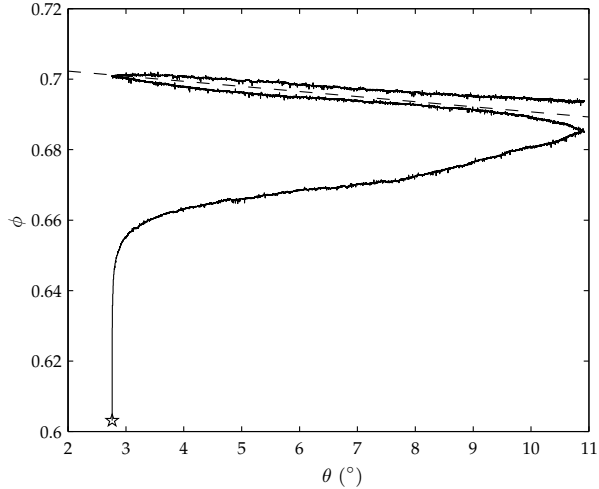}}
	\caption{Compaction under variable shear angle. $\theta$ is 
increased linearly from $2.7\,^\circ$ to $10.7\,^\circ$ then 
decreased to $2.7\,^\circ$ and finally increased again up to 
$10.7\,^\circ$. Rate of increase or decrease is $\pm 2.7\,10^{-4}\ ^\circ/$cycle. The dashed line has a slope equals to $-\alpha$.}
	\label{fig51}
\end{figure}

At this stage, it is interesting to compare the results presented 
here about the compaction of a granular packing under cyclic 
shear with the results obtained previously in the case of a vertically 
vibrated packing.
In order to do so, we have carried out the annealing experiment 
analog to the 
one described in \cite{nowak98} for the tap excitation: the amplitude 
of shear is 
continuously increased then decreased and increased again. The 
evolution of the volume fraction is plotted in Fig. \ref{fig51}. As 
observed in the tap 
experiment, one get an irreversible branch and a reversible one. 

However, in our case, the reversible branch is simply a straight 
line parallel to -$\alpha \theta$. 
In the light of the above interpretation about the response of the 
system to sudden 
shear changes,  this behavior can be explained as follows:
slowly increasing the shear amplitude yields
a quasi-saturated state, where $\phi_{slow}(n)$ does no longer vary 
with $n$. By then changing the angle $\theta$, one explores the linear 
variation of $\phi$ with $\theta$. The 
slight shift between the decreasing and increasing branches in the 
reversible regime comes from the slight 
evolution of $\phi_{slow}(n)$.  Hence this shift increases (resp. decreases) when decreasing 
(resp. increasing) the rate of change in angle. 

There thus exist several similarities between the two modes of 
compaction. However, some differences 
have to be noted.  First, cyclic 
shear induces the crystallisation of an assembly of monodispersed 
beads, whereas such order is not reported in the experiments of 
vertical taps. The small 
aspect ratio ($10$ particles in a cross-section) of the tap 
experiment could perhaps inhibit the crystalisation.
A second difference is the amplitude of the fluctuations. According 
to \cite{nowak98} vertical 
taps induce fluctuations from one cycle to another of the order of 
$0.8\%$ of the mean 
volume fraction, whereas in our shear experiment the variation of 
volume fraction is less than $0.1\%$. Periodic shear is perhaps a 
less violent way of exploring the different packing configurations.

\section{Conclusion}
\label{s:conclusion}

In this paper we have presented experiments about the compaction of a 
granular media under cyclic shear. First we have put in evidence that 
this 
method of compaction leads to crystals when the material is 
made of monodispersed particles. By studying the response of the 
system to sudden change in shear amplitude we have shown that 
discontinuities in 
volume fraction are observed, with amplitudes  proportional to the 
angle change. The evolution of the packing is thus composed of 
two dynamics: a slow, continuous and monotonous compaction and a 
rapid 
evolution occurring during the shear jump. 
This observation suggests that the rearrangement of the particles in 
the 
packing associated to the two dynamics are of different nature. 
However, the next important step would be to experimentally relate 
the compaction 
process with the evolution of the internal structure of the packing.

\begin{acknowledgement}
This work has benefited from fruitful 
discussions with R. Monasson, A. Lemaitre and S. Luding.  It  
 would not have been possible without the technical 
assistance of F. Ratouchniak. 
\end{acknowledgement}


\end{document}